\documentclass[superscriptaddress,twocolumn,aps]{revtex4}
\usepackage{graphicx}
\usepackage{dcolumn}
\usepackage{bm}
\usepackage{float}

\begin{document}

\title{Single atom adhesion in optimized gold nanojunctions}
\author{M.L. Trouwborst}
\affiliation{Physics of Nanodevices,
Zernike institute for advanced materials, Rijksuniversiteit
Groningen, Nijenborgh 4, 9747 AG Groningen, The Netherlands}
\author{E.H. Huisman}
\affiliation{Physics of Nanodevices, Zernike institute for advanced
materials, Rijksuniversiteit Groningen, Nijenborgh 4, 9747 AG
Groningen, The Netherlands}
\author{F.L. Bakker}
\affiliation{Physics of Nanodevices, Zernike institute for advanced
materials, Rijksuniversiteit Groningen, Nijenborgh 4, 9747 AG
Groningen, The Netherlands}
\author{S.J. van der Molen}
\affiliation{Kamerlingh Onnes Laboratorium,  Leiden University,
P.O. Box 9504, 2300 RA Leiden, The Netherlands}
\author{B.J. van Wees}
\affiliation{Physics of Nanodevices, Zernike institute for advanced
materials, Rijksuniversiteit Groningen, Nijenborgh 4, 9747 AG
Groningen, The Netherlands}
\date{\today}
\begin{abstract}
We study the interaction between single apex atoms in a metallic
contact, using the break junction geometry. By carefully 'training'
our samples, we create stable junctions in which no further atomic
reorganization takes place. This allows us to study the relation
between the so-called jump out of contact (from contact to
tunnelling regime) and jump to contact (from tunnelling to contact
regime) in detail. Our data can be fully understood within a
relatively simple elastic model, where the elasticity k of the
electrodes is the only free parameter. We find $5<k<32$ N/m.
Furthermore, the interaction between the two apex atoms on both
electrodes, observed as a change of slope in the tunnelling regime,
is accounted for by the same model.

\end{abstract}
\maketitle
Many macroscopic phenomena find their origin on the nanoscale,
since they are ultimately due to the interaction between single
atoms. A good example is formed by friction and wear, which have
been studied for centuries, but still inspire fascinating
research. For example, several groups have recently explored
methods to minimize friction in nanoelectromechanical systems
(NEMS), where no liquid lubricants can be applied\cite{friction}.
In this Letter, we focus on the ultimate miniaturization of the
problem and investigate adhesion and elasticity on the atomic
scale.\\

\begin{figure}[ht]
\begin{center}
\includegraphics[width=6cm, angle=-90]{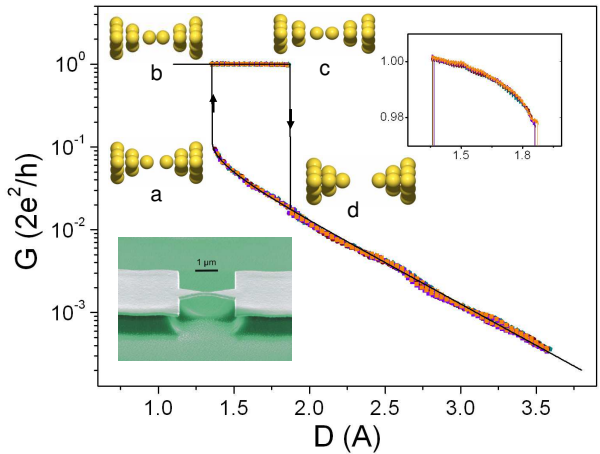}
\end{center}
\caption{Points: conductance G vs. distance D for four successive
$G_{0}$-loops ($V_{bias}=50$ mV). The jump to contact occurs at
D=1.5 ${\AA}$; the jump out of contact at D=2.0 ${\AA}$. Black line:
fit to model in Fig. \ref{model} (k=15.7 N/m). For the other
parameters we use literature values: $F_0=1.5nN$, $d=2.5 {\AA}$, and
$E_b=0.7eV$. The work function, $\phi=5eV$, was measured
independently. Inset graph: zoom of G vs. D in contact regime
(linear scale). Picture: scanning electron micrograph of a
lithographic break junction.} \label{G0loop}
\end{figure}

The interaction between single atoms can be studied by carefully
extending a notched metallic wire, while monitoring its
conductance $G$. As the wire is thinned out, $G$ decreases, until
its value is dominated by a few atoms forming a constriction
\cite{review}. When pulling is continued, an abrupt rupture of the
wire is observed (see Fig. \ref{G0loop}, point c to d). Upon
closing the contacts, a second jump occurs for many metals,
including gold (see Fig. \ref{G0loop}, point a to b )
\cite{gimzewski, krans}. These jumps are known as the 'jump out of
contact' (JOC) and 'jump to contact' (JC), respectively. By
carefully studying these discontinuities, one can in principle
obtain detailed information on the adhesion forces between two
single atoms. However, the details of the hysteretic loop in Fig.
\ref{G0loop} are still not fully understood. In fact, no relation
between the JC and JOC has been observed so far. The reason for
this is that the breaking process is generally accompanied by
plastic deformation, i.e., the atoms first reorganize before
rupture \cite{Rubio}. Therefore, during closing and opening, the
electrodes have a different atomic configuration. The most
intriguing example of plastic deformation is the formation of
atomic chains prior to breaking \cite{chains}.\\
Here, we explore junctions in which no plastic deformation occurs
during breaking and making of the single atom contact. This is
achieved by properly 'training' each device first. Figure
\ref{G0loop} displays G during the opening and closing of a
'trained' Au wire. Note that the single gold atom conductance is
characterized by a value close to the conductance quantum $G_0 =
2e^2/h$. As can be seen from the absence of conductance steps in
the contact regime, no atomic reorganization takes place.
Moreover, the curves in Fig. \ref{G0loop} are perfectly
reproducible for tens of subsequent runs. Being able to exclude
plasticity, we infer that the two jumps (JC and JOC) in Fig.
\ref{G0loop} are related to the adhesive forces between the single
atoms forming the junction. The remarkable reproducibility of the
'trained' junctions allows us to test a generic potential energy
model. In fact, we show that the whole making and breaking process
can be fitted by a single fit parameter: the elasticity of the electrodes.\\

\begin{figure}[h]
\begin{center}
\includegraphics[width=9cm]{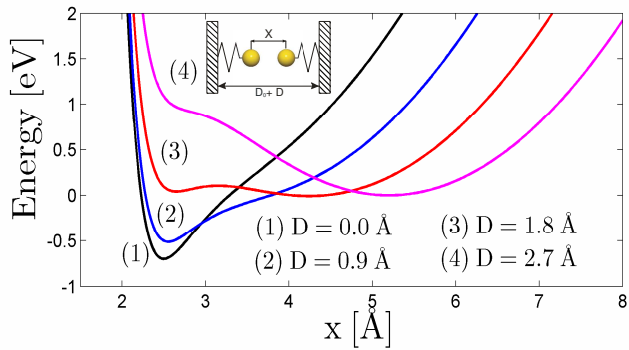}
\end{center}
\caption{Total energy as a function of the inter-atomic distance x
of a gold dimer, for different electrode separations D (see inset;
$D_0$ denotes an offset distance). Two contributions are included:
one due to the springs (spring constant k) and one due to the dimer
(described by the 'universal' binding curve). Depending on D, the
total energy may have two minima. For the atom to jump in and out of
contact, a barrier has to be overcome. Hence, opening and closing
occurs at different positions, explaining the hysteresis in $G_{0}$
loops. Note: same parameters as in Fig. \ref{G0loop}.} \label{model}
\end{figure}

For our study, it is crucial to minimize drift and vibrations in
the electrodes during the measurements. This is achieved by using
a mechanically controllable break junction (MCBJ) at 4.2 K
\cite{mcbj}. The wires are patterned with electron beam
lithography, after which we evaporate 1 nm of Cr and 120 nm Au at
$2*10^{-7}$ mbar. Finally, the area below the wire is etched with
a $CF_{4}/O_{2}$ plasma, to create a free hanging gold wire [inset
in Fig. \ref{G0loop}]. The MCBJ is cooled down to 4.2 K and broken
by bending the substrate. Thus, clean and stable gold electrodes
are obtained in cryogenic vacuum. Opening and closing of the
junction is done with an effective speed of 0.5 $\AA/s$, while the
conductance, at a 50 mV bias, is monitored. Measurements at 1-300
mV gave similar results\cite{voltage1}. We emphasize the
impressive stability of the electrodes, resulting in a drift below
0.3 pm/h. Our break junctions are calibrated using Gundlach
oscillations \cite{calibratie1, calibratie2}. We find an
attenuation factor $r=(5.4\pm0.6) \cdot 10^{-5}$ \cite{vrouwe,
Trouwborst}, and a work function $\phi=5 eV$.\\

The details of our 'training' method are as follows. When closing
the electrodes, we stop immediately as soon as the electrodes are in
contact, i.e. at $G\approx G_0$, preventing further disorder.
Subsequently, we break the wire, extend it 1-2 $\AA$ into the
tunnelling regime and close it again until the jump to contact.
Repeating this procedure rearranges and orders the tip atoms. In
this way, the atoms are able to probe (energetically and spatially)
different positions, allowing them to find the most stable
configuration. Remarkably, after typically $>$10 sweeps, JC and JOC
occur at two \emph{exactly} reproducible positions. In Fig.
\ref{G0loop}, four subsequent traces are shown with perfect
repeatability. In fact, the maximum variation in the closing and
opening points was less than 5 pm over 50 sweeps. Although these
loops (which we call '$G_{0}$-loops') have already been observed by
other groups \cite{krans,metallic adhesion}, we are the first to
optimize the training method to investigate JC and JOC in
well-defined geometries. We have measured 734 different $G_0$ loops,
on 8 different samples, to study the variation in the contacts. To
obtain a new $G_{0}$-loop, we first rearrange a contact by closing
up to $>10G_{0}$, before training the contact for a different
$G_{0}$-loop. For each $G_{0}$-loop, we automatically record the
four conductance values $G_a$, $G_b$, $G_c$ en $G_d$ (at points
a,b,c and d in Fig. \ref{G0loop}, respectively). The fact that $85
\%$ of the conductance values $G_c$ is above $0.9 G_0$ emphasizes
the good definition of our junctions. Recently, Untiedt \textit{et
al.} studied JC (no 'training' method was employed) \cite{untiedt}.
A statistical analysis was made of many closing traces and
correlations were found between the conductance values just before
and just after JC ($G_a$ and $G_b$). For gold, they observed maxima
in density plots, for $G_b$ values below $G_0$ and around 1.6 $G_0$.
Conductances below 1 $G_0$ were attributed to a dimer configuration, whereas higher conductances were related to monomer and double bond configurations. We have also observed the peak around
1.6 $G_0$ for untrained junctions. However, upon breaking G dropped
in steps to lower conductances, making it impossible to create
stable '$G_0$-loops' for this configuration. Furthermore, for our
trained contacts, more than 80 $\%$ of our $G_b$ values have a conductance below 1.02 $G_0$. Therefore, we conclude that training of the contacts results predominantly in the dimer configuration, as sketched in Fig.
\ref{G0loop}. Also from molecular dynamics simulations, dimers are
expected to be the most stable geometry
\cite{Dreher,untiedt}.\\

Since for the trained contacts all plastic deformation is removed,
we can employ an elastic model to describe the hysteretic loop in
Fig. \ref{G0loop}. The basic configuration is shown in the inset
of Fig. \ref{model}, where a dimer is depicted in between two
elastic electrodes. To describe the force between the two atoms,
we use the so called 'universal' binding curve, which is given by
\cite{Rose,Bahn,Krans}

\begin{equation}\label{universal}
    E(x)=-\alpha(x-x_{0})e^{-\beta(x-x_{0})}
\end{equation}

where x is the interatomic distance. The parameters $\alpha$,
$\beta$ and $x_0$ are related to the equilibrium bond distance
$d=x_0+1/\beta$, the binding energy $E_b=-\alpha/\beta e$ and the
slope at the inflection point $F_0=\alpha/e^2$, with $e=2.718$.
These values are known from literature, i.e., $d=2.5 \pm 0.2\AA$
\cite{calibratie2}, $F_0=1.5 \pm 0.3nN$ for the break force
\cite{Rubio-bollinger} and $E_b=0.7 \pm 0.2 eV$ for the binding
energy \cite{metallic adhesion}. For the bonding energy of the
dimer to the rest of the electrodes (the "banks"), we take the
potential energy of a spring ($ku^{2}/2$). Hence, we are left with
only one free parameter, the spring constant k, which is directly
related to the hysteresis of the $G_{0}$-loop. In Fig.
\ref{model}, the total energy is plotted versus the interatomic
distance $x$, for different electrode distances D. Depending on D,
two minima are present. When starting with a closed contact in
equilibrium, i.e. D=0 $\AA$, the equilibrium interatomic distance
equals $x_{eq}=2.5 \AA$ (minimum of curve 1). As the electrodes
are pulled apart, e.g., by 1.8 $\AA$ (curve 3), the two atoms are
separated by only 0.2 $\AA$. The rest of the displacement is
invested in stretching the spring. Increasing D further (towards
curve 4), the first minimum disappears and the system jumps to the
second minimum (at $x_{eq} \approx 5.3 \AA$); this is JOC. The
atoms of the dimer get separated by typically 2-3 $\AA$. Upon
closing the junction, a linear relation between $x_{eq}$ and D is
initially seen. However, at small separations, the spring
stretches somewhat due to the attractive forces of the opposing
atoms, i.e. $x_{eq}$ moves faster than D\cite{metallic adhesion,
hofer}. This gives a deviation from exponential tunnelling, as
observed in Fig. \ref{G0loop}. The effect is maximal just before
JC, which takes place in between curves 3 and 2, and covers a
distance of $\approx 1 \AA$. To apply our model to the tunnelling
part of the $G_{0}$-loops, we assume that the work function,
$\phi$, does not depend on $x_{eq}$, so that $G\propto
exp(-2x_{eq}\sqrt{2m\phi}/\hbar)$. Taking $G=G_0$ for $D=0$, we
have the tools to fit the data in Fig. \ref{G0loop}. The
corresponding trace is shown in Fig. \ref{G0loop}. It gives a
perfect fit, not only to the exact position of JC and JOC, but
also to the deviation from exponential tunnelling. For this
$G_{0}$-loop, the fit parameter k assumes a value $k=15.7$ N/m.\\

\begin{figure}[!h]
\begin{center}
\includegraphics[width=9cm]{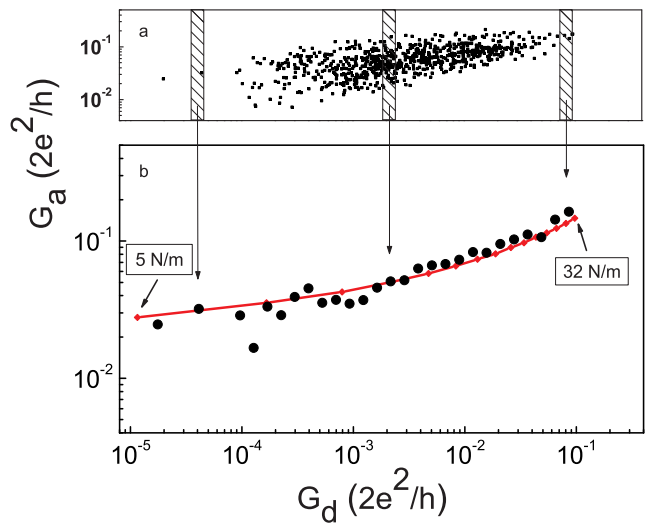}
\end{center}
\caption{(a) Conductance G just before JC ($G_a$) vs. G just after
JOC ($G_d$) for 734 different $G_{0}$-loops. (b) Average $G_a$ as a
function of $G_d$. Averaging is done within regular bins of $G_d$,
as indicated in a). The line is a fit to the model, assuming a
varying spring constant $5 < k <32$ N/m. The other parameters are
the same as in Fig. \ref{G0loop} and \ref{model}.} \label{GaGd}
\end{figure}

In total, we studied 734 $G_{0}$-loops, which could all be fitted
with the model. In Fig. \ref{GaGd}, the conductance $G_a$ (just
before JC) is plotted versus $G_d$ (just after JOC), for all
$G_{0}$-loops measured. Remarkably, the graph shows a relation
between these points. This is especially visible in Fig.
\ref{GaGd}b, where the average $G_a$ is displayed versus $G_d$. We
compare these data points to our model. By varying the elasticity k
from 5-32 N/m, with all other parameters fixed to literature values,
one obtains the line drawn in Fig. \ref{GaGd}. Clearly, the data
points are well described by the model, using only k as a variable.
Note that the range of k-values is in agreement with Ref.
\cite{metallic adhesion}. Moreover, $>$ 90 $\%$ of our k values
(per electrode) are in the range 7-26 N/m. This variation is
substantially smaller than the spread found in Ref \cite{metallic
adhesion}. We relate the spread in k to the following phenomena.
First, breaking a gold wire is only possible along certain crystal
orientations ([111], [100] and [110]) \cite{rodrigues}. For each
orientation, the apex atom is bonded differently to the second layer
of the electrode. In fact, the apex atom has 3, 4 and 5 nearest
neighbors for the Au [111], [100] and [110] direction, respectively.
This is expected to have substantial influence on the elasticity of
the electrode. However, this picture is not yet complete. The work
by Olesen implies that k is only partly determined by the nearest
neighbors of the apex atom \cite{olesen}. In fact, the most
significant contribution to k arises from elastic displacements in
the rest of the metal tips. Hence, k is related to the precise
structure of the electrodes on a larger scale, which varies with
each $G_0$ loop. Every time we close the junction up to 10 $G_0$, we
most likely introduce defects in the atomic layers further into the
contact \cite{Yanson}, which influence the elasticity of the
electrodes. A final source of variation is the so-called "lateral
approach" of the electrodes. For the model in Fig. \ref{model}, we
assumed that the apex atoms are perfectly aligned. However, small
misalignments (0-1 ${\AA}$) may occur in reality. We extended the 1D
model to a 2D model by assuming springs in both x and y directions.
This yields a variation in k of up to 10 $\%$. We stress that the
other parameters in our model cannot explain the data in Fig.
\ref{GaGd}. Only the spring constant gives a relation along the
direction shown. The variation in the vertical direction (Fig.
\ref{GaGd}a), however, can be explained by small deviations in
$E_b$. As indeed shown by Ref. \cite{Bahn}, $E_b$ is sensitive to
the local atomic configuration. We find that a spread of 25 $\%$ in
$E_b$ explains the variation in $G_a$. Such a spread is consistent
with the results of Ref. \cite{metallic adhesion}. We conclude that
our relatively simple model does not only explain the occurrence of
JC and JOC, but also their relationship. Finally, we emphasize that
without 'training', atomic reorganization upon opening and closing
destroys the
interdependence between $G_a$ and $G_d$.\\
Remarkably, our model fits the data in Figs. \ref{G0loop} and
\ref{GaGd} very well, despite the fact that we assume a constant
barrier height $\phi$ for all $x_{eq}$. This contrasts
computations by Lang which predict a strong decrease of the
apparent barrier at electrode distances $<4 \AA$ \cite{lang}. Such
barrier lowering would primarily be due to the image forces and
local effects related to the electric charges on the electrodes.
Olesen \textit{et al} carefully examined the tunnel curves of Ni,
Pt and Au by scanning tunnelling microscopy (STM) and found no
deviations from exponential behavior. This was explained by
assuming that barrier lowering is exactly cancelled by adhesion
between tip and sample\cite{olesen}. The fact that we can fit our
$G_{0}$-loops using an elastic model only, shows that the
influence of image forces in break junctions (which feature
two sharp tips) is relatively small.\\

\begin{figure}[!h]
\begin{center}
\includegraphics[width=9cm]{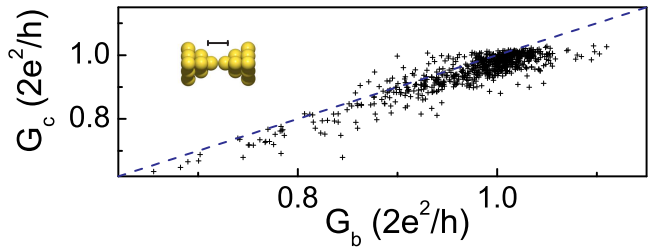}
\end{center}
\caption{(a) Conductance just before JOC ($G_c$) vs. G just after JC
($G_b$) for 734 different $G_{0}$-loops. The dashed line represents
$G_c=G_b$. Inset: schematic junction; scale bar: barrier length for
tunnelling.} \label{GbGc}
\end{figure}

Having described three quarters of the $G_{0}$-loop (JOC,
tunnelling and JC), we focus on the contact regime. In Fig.
\ref{GbGc}, $G_c$ is plotted versus $G_b$. In all cases, we have
$G_c \leq G_0$, as expected for the conductance of a single atom
contact. In contrast, the values of $G_b$ go well above 1 $G_0$.
On average, $G_b$ is 0.025 $G_0$ higher than $G_c$. This is
exactly as expected, if a second conductance channel is
considered. As calculated by Ref. \cite{Dreher}, the small
distance between the second layers of the two electrodes allows
for tunnelling with a conductance of at most 0.03 $G_0$. When
stretching the contact, however, the tunnel gap increases and the
transmission of the second channel tends to zero. This is why it
is not visible in $G_c$, where the contact is fully extended.\\

In summary, we have created highly ordered gold electrodes, using
a "training" method. Upon opening and limited closing, our
junctions show no plastic deformation, allowing us to study the
jump out of contact, tunnelling curves and jump to contact in
detail. Individual breaking and making loops can be perfectly
fitted with an elastic model, having the spring constant of the
electrodes, k, as the only free parameter. Hence, by suppressing
plastic deformation effects, we are able to measure and model
adhesion on the single atomic level.

This work was financed by the Nederlandse Organisatie voor
Wetenschappelijk onderzoek, NWO, via a Pionier grant, and the
Zernike Institute of Advanced Materials. We thank Simon Vrouwe,
Siemon Bakker and Bernard Wolfs for technical support and Jan van
Ruitenbeek for useful discussions.

\end{document}